\documentclass{emulateapj}
\usepackage{color}
\usepackage{graphicx}
\usepackage{amsfonts,amsmath,amssymb}
\usepackage{natbib}
\usepackage{hyperref}
\hypersetup{colorlinks=false,pdfborder={0 0 0}}
\usepackage[utf8]{inputenc}
\usepackage[english]{babel}

\newcommand{\atomH}{\ensuremath{\mathrm{H}\, \mathrm{I}}}
\newcommand{\molH}{\ensuremath{\mathrm{H}_2}}
\newcommand{\CO}{\ensuremath{\mathrm{CO}}}
\newcommand{\glon}{\ensuremath{\ell}}
\newcommand{\glat}{\ensuremath{\mathrm{b}}}
\newcommand{\RGC}{\ensuremath{R_{\mathrm{GC}}}}
\newcommand{\vlos}{\ensuremath{v_{\mathrm{los}}}}
\newcommand{\Reid}{{Reid14}}

\slugcomment{To appear in The Astronomical Journal (AJ)}

\shorttitle{Kinetic Tomography I: A Method for Mapping the Milky Way's Interstellar Medium in Four Dimensions}
\shortauthors{Kirill Tchernyshyov}

\begin{document}

\title{Kinetic Tomography I: A Method for Mapping the Milky Way's Interstellar Medium in Four Dimensions}

\author{Kirill Tchernyshyov\affil{The Johns Hopkins University} J.E.G. Peek\affil{Space Telescope Science Institute}}

\begin{abstract}

We have developed a method for deriving the distribution of the Milky Way's interstellar medium as a function of longitude, latitude, distance and line-of-sight velocity.
This method takes as input maps of reddening as a function of longitude, latitude, and distance and maps of line emission as a function of longitude, latitude, and line-of-sight velocity.
We have applied this method to datasets covering much of the Galactic plane.
The output of this method correctly reproduces the line-of-sight velocities of high-mass star forming regions with known distances from \citet{2014ApJ...783..130R} and qualitatively agrees with results from the Milky Way kinematics literature. 
These maps will be useful for measuring flows of gas around the Milky Way's spiral arms and into and out of giant molecular clouds.

\end{abstract}

\keywords{ISM: kinematics and dynamics, methods: statistical}

\section{Introduction}
Many open problems in star formation, molecular cloud evolution, and galaxy-scale gas dynamics remain open because it has not been possible to measure the most useful quantities for resolving them -- the 3D gas velocity vector and 3D gas density over an extended area of sky. A measurement of these fields would allow us to solve the continuity equation \citep{euler1757principes}, and derive the rate at which density is changing across the Galaxy over a range of physical scales. 

The formation of giant molecular clouds (GMCs), for instance, is in part a matter of collecting a large mass in a small volume. 
By looking for sites at which gas flows are converging, it may be possible to find currently forming GMCs.
Conversely, one could look for diverging flows to detect and characterize the feedback-driven disruption of GMCs.
With sufficient spatial and velocity resolution, it would be possible to distinguish between the many theories of exactly how the necessary mass is accumulated and converted to cold, molecular gas. 
These theories make different predictions for the properties of the required converging flows.
For example, some theories assume the converging flows consist mostly of neutral hydrogen (\atomH) and invoke different sorts of instabilities at the collision interface of the flows to explain how this gas is rapidly converted to cold molecular hydrogen (\molH) \citep[e.g. ][]{Heitsch06,Clark:2012bq,2014ApJ...790...37C}.
Other theories assume that GMCs form from the collision and agglomeration of smaller molecular cloudlets \citep[e.g. ][]{Roberts:1987eb,Dobbs:2008ez,Tasker:2009gc}. 
One could directly distinguish between these two possible modes of GMC formation by determining whether gas that is converging on forming GMCs is predominantly neutral or molecular.
Without the ability to measure the velocity field as a function of all three spatial dimensions, it is difficult to even determine where converging flows are present.

The origin of the converging flows invoked above is also a matter of active interest.
In some theories, the self-gravity of a modest overdensity can be sufficient to induce collapse \citep{Kim:2002da,VazquezSemadeni:2007cj,2012MNRAS.425.2157D}.
Others invoke spatially coherent flows driven by feedback from star formation \citep{Fujimoto:2014kh} or perturbations in the Galactic potential such as spiral arms \citep{Roberts:1972bp,Bonnell:2006hn}.
Strong, shocked flows driven by spiral arms have been seen in strongly tidally interacting two-arm spiral galaxies in the nearby universe \citep{Visser:1980ud,Visser:1980vc,Shetty_2007} using $\CO$ and $\atomH$ observations.
Unfortunately, the resolution in $\atomH$ required to map these spiral shocks beyond very nearby galaxies with extreme two-armed spiral structure, is not observationally feasible \citep{Visser:1980ud}.

The nearest spiral galaxy is, of course, our own Milky Way. 
Studying the kinematics of the Milky Way replaces the problem of insufficient sensitivity and spatial resolution with the problem of confusion --- from our vantage point, it is difficult to determine how the ISM is moving as a function of 3D position.
In particular, it is essentially impossible to obtain the transverse velocity field of the ISM. Transverse velocities can only be derived from proper motions, which are difficult to measure for the diffuse and continuous ISM.
For many of the open problems we have outlined, even a measurement of the line-of-sight (radial) velocity as a function of 3D position (the 1-velocity field) would represent a significant step forward. 
While only having one component of the velocity field does make it difficult to empirically measure the  rate of inflow and outflow of matter relative to any given structure, it should still be possible to empirically measure this rate statistically across a sample of structures. 

No single observable can simultaneously measure the density and line-of-sight velocity at each location in the ISM, a construct we call the position-position-distance-velocity (PPDV) 4-cube. 
Instead, we have a number of observables of the ISM, each of which can be considered a (noisy and biased) projection of this PPDV 4-cube.
We dub the process of reconstructing this PPDV 4-cube from some set of observations ``Kinetic Tomography'' (KT). There are a number of existing kinds of data we can use. The first are classical radio observations of the ISM in both diffuse tracers (e.g. $\atomH$) and denser tracers (e.g. $\CO$). These are position-position-velocity (PPV) 3-cubes, a specific projection of the PPDV 4-cube, and have been classically used to infer the density of the ISM in 3-space assuming a Galactic rotation curve \citep[e.g.][ and references therein]{Levine_2006}. Another observable is the position-position-distance (PPD) reddening 3-cubes generated by examining the photometry of large numbers of stars and performing inference on the intervening dusty ISM. There has been dramatic progress in this field \citep{Marshall_2006,Lallement_2014,Green_2015}, which has been crucial for allowing this investigation. 
Interstellar absorption lines toward stars also represent a projection of the PPDV 4-cube and can simultaneously contain distance, column density, and velocity information about the intervening matter \citep{Welsh10,2015ApJ...798...35Z,2015MmSAI..86..521Z}.

There have been some attempts to construct maps or point estimates of $\vlos$ as a function of distance. 
Most such attempts have focused on individual spiral arms and used models of ISM flows around the spiral arms to directly invert PPV 3-cubes (e.g  \citealt{1972A&A....16..118S,Foster_2006}).
\citet{Reid_2016} have developed a different approach, which combines probability distributions from the standard kinematic distance, various geometric hints, and possible associations of emitting gas with structures that have parallax-based distance measurements into a combined syncretic probability distribution for the distance. 
Neither of these approaches uses the information available in reddening-based PPD 3-cubes. 

In the method we describe in this work, we use large-area CO and HI PPV 3-cubes and the \citet{Green_2015} (henceforth GSF) PPD 3-cube to reconstruct the ISM PPDV 4-cube. We perform a restricted version of the full tomographic reconstruction in which we assume each parcel of gas in the PPD 3-cube is assigned a single central line-of-sight velocity with some line-of-sight velocity width. This can be considered an inversion of the usual kinematic distance method, in which a line-of-sight velocity is converted to a distance using a Galactic rotation curve. Here we map distance along a sightline to velocity and allow deviations from a rotation curve in order to better match the distribution of matter in the PPV 3-cube.

In this work, we describe our method, the map it produces, and our evaluation of this map's accuracy and precision.
Subsequent papers in this series will analyze this map to gain insights into Milky Way kinematics and their connection to star formation. 
In Section \ref{sec:data}, we describe the datasets we use to make and evaluate our PPDV map.
In Section \ref{sec:KT}, we give a detailed explanation of the PPDV mapping technique and quantitatively demonstrate the accuracy of the technique's results.
In Section \ref{sec:discussion}, we discuss the broader accuracy and applicability of the technique.
We conclude in Section \ref{sec:conclusion}.

\begin{figure*}[t]
\begin{center}
\includegraphics[width=1\linewidth]{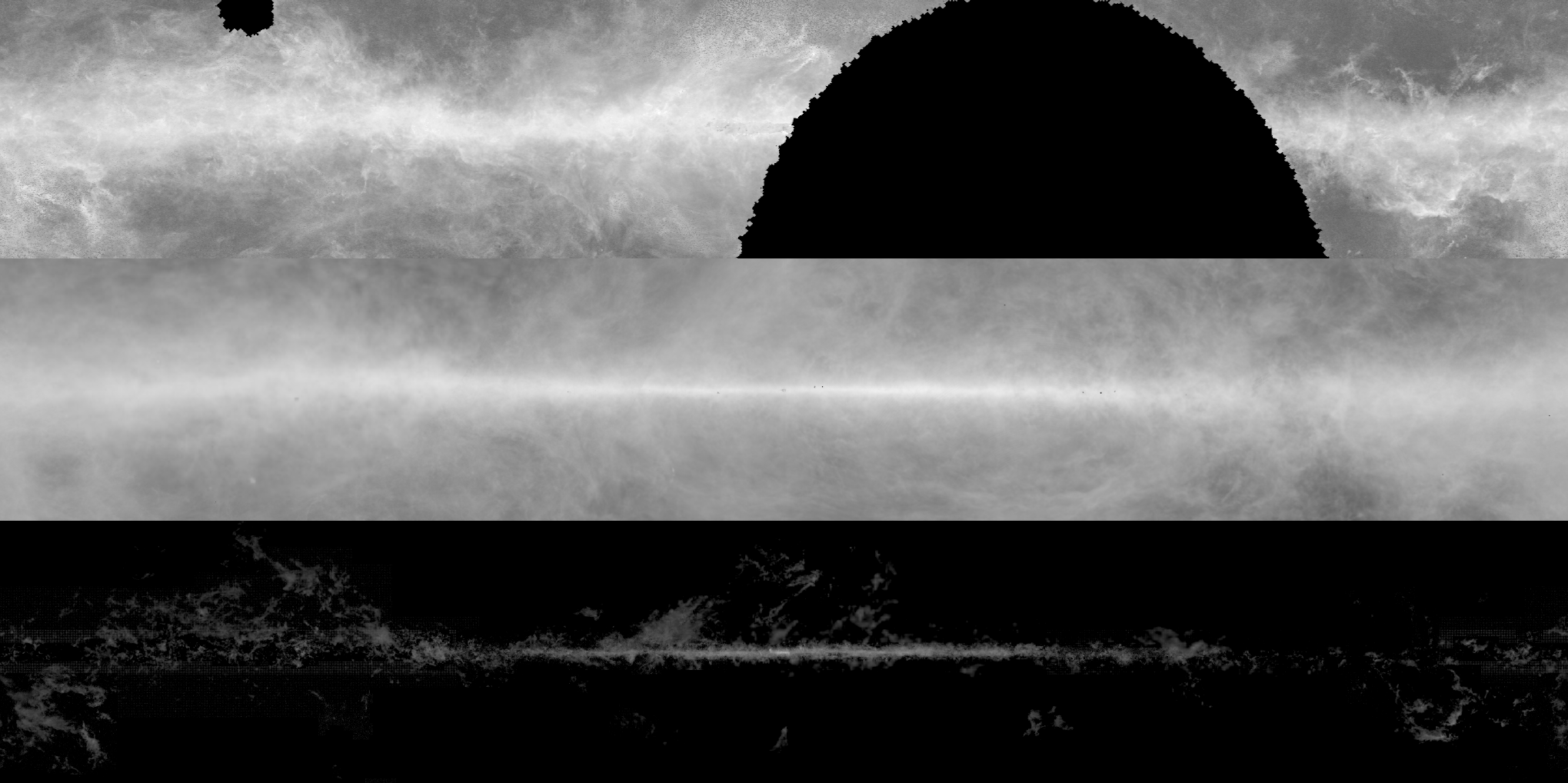}
\caption{{The interstellar medium in the Galactic plane, shown in three tracers: reddening (top), $\atomH$ (middle), and $\CO$ (bottom). Each shows the total column measured by each tracer, with a log intensity stretch. The regions are 360 degrees in Galactic longitude wide, centered on Galactic Center, and 60 degrees in Galactic latitude tall, centered on the Galactic plane.
}}
\end{center}
\end{figure*}

\section{Data}
\label{sec:data}
\subsection{HI and CO data}

Radio emission lines of $\atomH$ and $\CO$ trace the two dominant constituents of the Galactic ISM, atomic and molecular gas. 
Ionized phases of the ISM do not contribute significantly to the column density and will therefore contribute negligibly to the extinction measured in GSF. 
21-cm line emission from the hyperfine transition of $\atomH$ is usually optically thin and its integral is an excellent tracer of $\atomH$ column:
\begin{equation}\label{XHI}
{\rm N(\atomH)} = 1.8 \times 10^{18} {\rm\, cm^{-2}} \frac{ \int T_B {\, \rm d}v}{\rm \, K~km~s^{-1}}.
\end{equation}
When the 21-cm line becomes optically thick, Equation \ref{XHI} will underestimate the $\atomH$ column. 
However, this mostly happens in $\molH$-dominated regions \citep{Goldsmith_2007}.

We trace molecular gas using the 115 GHz 1-0 rotational transition of $\CO$. 
The integral of this emission line can be converted to a $\molH$ column density using the conversion factor \citep{Bolatto_2013}

\begin{equation}\label{XCO}
X_{CO} = 2.0 \times 10^{20} {\rm\, cm^{-2}} \frac{ \int T_B {\, \rm d}v}{\rm \, K~km~s^{-1}}.
\end{equation}

This conversion factor has a number of known weaknesses stemming from complex excitation and opacity effects and real variation in the relative population of $\CO$ and $\molH$ molecules. 
We will address the impacts of these weaknesses in \S \ref{sec:discussion-systematics}. 

For our $\CO$ data, we use the interpolated whole-Galaxy PPV 3-cube provided by \citet{Dame_2001}. 
The 3-cube covers the full range in $\glon$, $-30^\circ$ to $+30^\circ$ in $\glat$, and $-320{\rm \,km~s} ^{-1}$ to $+320 \rm{~km~s}^{-1}$ in $V_{LSR}$ and has a velocity resolution of $1.3 {\rm \, km~s}^{-1}$.
We find that the native resolution and PPV extent of these data are appropriate for our investigation and retain their exact pixelization for our other datasets.

The $\CO$ 3-cube contains single-pixel artifacts in both emission and absorption. 
To remove these artifacts, we apply a plus-shaped median smoothing kernel to each velocity channel. 
This kernel is three pixels wide along the $\glon$ and $\glat$ directions.
This filtering procedure changes the total amount of CO emission by about 5\% over the entire Galaxy.

For $\atomH$ data, we use a combination of three large-area Galactic $\atomH$ surveys. 
South of declination 0$^\circ$, we use data from the 16$^\prime$ resolution GASS survey \citep{Kalberla_2010}; from declination 0$^\circ$ to 38$^\circ$ we use unpublished data from the 4$^\prime$ resolution GALFA-HI survey \citep{Peek_2011} Data Release 2; North of 38$^\circ$ we default to the 36$^\prime$ resolution LAB Survey \citep{Kalberla_2005}. 
We regrid these data onto the $7.5^\prime \times 7.5^\prime \times 1.3 {\rm \, km~s}^{-1}$ pixels of the \citet{Dame_2001} $\CO$ 3-cube.
We note that each of these surveys has potential pitfalls. LAB is quite low resolution compared to the other data sets, meaning small scale Galactic features may be lost. GALFA-HI does not have stray radiation correction applied and so may overestimate columns in low column density regimes. Small artifacts exist in all of these databases but are typically most pronounced in high-latitude fields. 
The possibility of differences in the accuracy of the KT solution between regions in which we use different $\atomH$ surveys is discussed in \S \ref{sec:discussion-catastrophic}.

The $\atomH$ and $\CO$ emission 3-cubes are converted to ${\rm N}(\atomH)$ and ${\rm N}(\molH)$ 3-cubes using Equations \ref{XHI} and \ref{XCO}. These two column density 3-cubes are then added to make a single $\mathrm{N_H}$ data 3-cube.

\subsection{Dust data}

Our extinction 3-cube is derived from the GSF reddening data. 
GSF use PanSTARRS photometry of 800 million stars to infer the cumulative reddening along the line of sight in 6.8$^\prime$ (NSIDE=512 HEALPix) pixels.
The distance axis of the GSF 3-cube is in steps of half a distance modulus, from 63 pc to 63 kpc. 
We regrid these data onto the \citet{Dame_2001} $\glon$-$\glat$ grid and difference them in distance to find the reddening between each distance step. 
This differential reddening is then converted to an ${\rm N_H}$ using the factor measured in \citet{Peek_2013}, 

\begin{equation}
\mathrm{N_H} = E\left(B-V\right) 7 \times 10^{21} \frac{\rm cm^{-2} }{\rm mag}. 
\end{equation}

We use a single reddening to hydrogen column density conversion factor. 
This is, in principle, incorrect due to variations in both the relation between the amount of reddening and the amount of dust (e.g. \citealt{2015A&A...582A..31P}) and the ratio of dust to gas (e.g. \citealt{1996ARA&A..34..279S,Jenkins:2009ke}).
We expect the latter to be the more important effect.
The combination of $\atomH$ and $\CO$ emission lines traces gas from at least three of the four cloud types in \citet{1996ARA&A..34..279S}, all of which have different depletion patterns and hence dust-to-gas ratios.
This change in depletion pattern corresponds to a change in dust-to-gas ratio of about two, which while not negligble is also unlikely to be the dominant systematic uncertainty in our analysis.
We discuss the impact of making this incorrect assumption in \S \ref{sec:discussion-systematics}.

\subsection{High-mass star forming region data}
\label{sec:data-HMSFR}
To check the accuracy of our method, we need measurements of the line-of-sight velocities and distances of clouds of gas. 
\citet{2014ApJ...783..130R}(henceforth \Reid{}) have measured the line-of-sight velocities, proper motions, and trigonometric parallaxes of water and methanol masers associated with 103 high-mass star forming regions (HMSFRs). 
Of these 103, 99 fall inside the footprint of the GSF reddening data. 
We adopt the line-of-sight velocities, velocity uncertainties, and parallaxes of these 99 HMSFRs as stated in Table 1 of \Reid{}. 

\begin{figure*}[t]
\begin{center}
\includegraphics[width=0.6\linewidth]{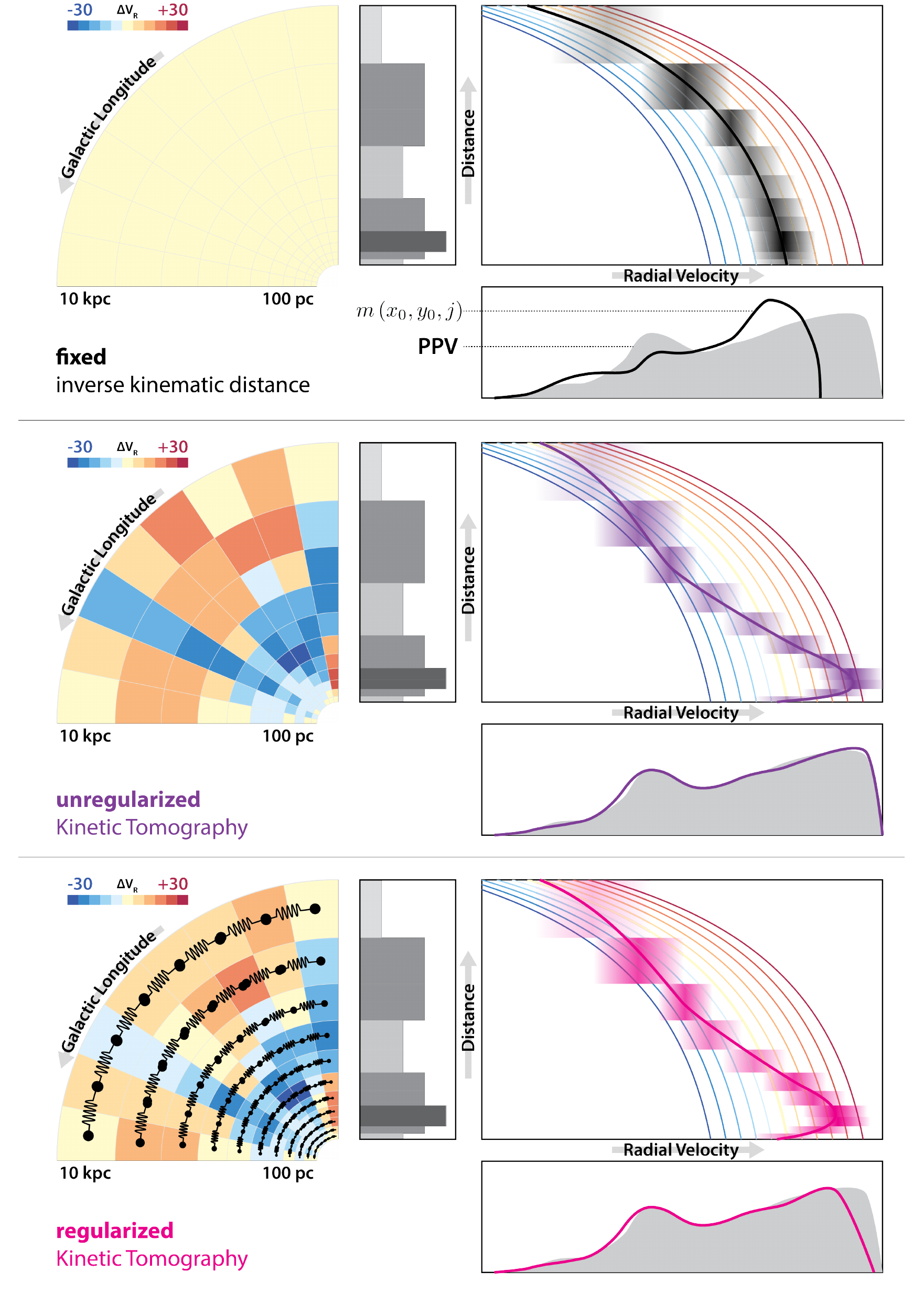}
\caption{{\label{fig:diagram} Diagrams explaining the inverse kinematic distance methods, unregularized Kinetic Tomography, and regularized Kinetic Tomography. In each diagram, we show a map of residuals from a flat rotation curve (left quarter-circle), mock PPD and PPV data along a single sightline $\left(x_0, y_0\right)$ (narrow vertical and horizontal panels), and the mean ($\vlos\left(x_0, y_0, i, j\right)$) and standard deviation ($\sigma_v\left(x_0, y_0, i, j\right)$) of the velocity along that sightline. The $\glon$, $\glat$, distance, and $\vlos$ axes are indexed by $x, y, i,$ and $j$; note that the residual map shows only a single latitude. The results of applying a fixed inverse kinematic distance method, unregularized Kinematic Tomography, and regularized Kinetic Tomography are shown in black, purple, and pink. Regularization is indicated in the bottom schematic with spring symbols. Our regularization method minimizes the velocity difference between voxels that are connected by spring symbols.}}
\end{center}
\end{figure*}

\section{Kinetic tomography}
\label{sec:KT}
We have developed a procedure for deriving the distribution of interstellar matter in PPDV space from measurements of its distribution in PPD and PPV space. 
The technical term for deriving a multi-dimensional distribution from lower-dimensional measurements is \emph{tomographic reconstruction}; in this context, lower-dimensional measurements are called \emph{projections}. 
Tomographic reconstruction from two projections is, in general, not possible -- there are many more independent variables than observational constraints. 
Our procedure for solving this specific case of the tomographic reconstruction problem uses simplifying assumptions about the structure of the ISM in PPDV space to reduce the number of independent variables.

For motivation, we first examine the assumptions behind a common technique for reconstructing a PPDV-space ISM distribution from PPV-space measurements alone. 
This is the widely known kinematic distance method, which maps line-of-sight velocities to distances based on an assumed rotation curve and Galactic geometry. 
While the kinematic distance method is usually presented as a way of converting a PPV-space distribution to a PPD-space, rather than PPDV-space, distribution, the two conversions are equivalent according to the kinematic distance method's underlying assumptions.
These assumptions can be combined into a single statement: (1) a location in PPD space can be assigned a single line-of-sight velocity (2) according to an assumed rotation curve and Galactic geometry. 
With these assumptions, it follows that knowing the PPD-space distribution of the ISM is equivalent to knowing its PPDV-space distribution. 

We use PPD measurements in addition to PPV measurements, allowing us to relax these assumptions. 
Our version: (1) a parcel of interstellar matter in PPD space can be assigned a \emph{Gaussian distribution} of line-of-sight velocities (2) whose center is \emph{within a fixed range of} the line-of-sight velocity predicted by an assumed rotation curve and Galactic geometry. 
Here, a ``parcel'' of interstellar matter refers to the contents of a single PPD 3-cube voxel. 
That is to say, we aim to assign a central line-of-sight velocity $\vlos$ and a line-of-sight Gaussian velocity width $\sigma_v$ to each voxel in the PPD 3-cube. 
From our assumptions, a description of the ISM in PPDV space consists of a description of its PPD-space distribution (the observed PPD 3-cube), a line-of-sight central velocity 3-cube in PPD space ($\vlos(\glon, \glat, d)$), and a line-of-sight velocity width 3-cube in PPD space ($\sigma_v (\glon, \glat, d)$). 
Thus, we have reduced our original problem of finding a PPDV 4-cube which is consistent with our observed PPD and PPV 3-cubes to finding a $\vlos(\glon, \glat, d)$ and $\sigma_v(\glon, \glat, d)$ pair consistent with our PPV observations.

\subsection{Formalism}
\label{sec:KT-method}
We are looking for a $\vlos(\glon, \glat, d)$ and $\sigma_v(\glon, \glat, d)$ pair that, when combined with the observed PPD 3-cube, best matches the observed PPV 3-cube. 
This is an optimization problem. 
An optimization problem needs an objective function, and an objective function needs a quantity to compare to the data. 
In our case, that quantity is a model PPV 3-cube.
To obtain a model PPV 3-cube, we produce a PPDV 4-cube and then integrate it along the distance axis. 

Suppose that we have a matter density $\rho(\glon, \glat, d)$ in PPD space and a $\vlos(\glon, \glat, d)$ and $\sigma_v(\glon, \glat, d)$ pair. 
The matter density at a point $(\glon, \glat, d, v)$ in PPDV space is 
\begin{equation}
 \label{eqn:ppdv_cont}
  \rho(\glon, \glat, d, v) = \frac{\rho(\glon, \glat, d)}{\sqrt{2 \pi \sigma_v(\glon, \glat, d)^2}} 
   \, \exp\left(- \frac{\left( v - \vlos(\glon, \glat, d) \right)^2}{2 \sigma_v(\glon, \glat, d)^2} \right),
\end{equation}
and the matter density at a point $(\glon, \glat, v)$ in PPV space is
\begin{equation}
  \label{eqn:ppv_cont}
  \rho(\glon, \glat, v) = \int_0^{d_{max}} \rho(\glon, \glat, d, v) \, {\rm d}d,
\end{equation}

where $d_{max}$ is the maximum distance to which $\rho(\glon, \glat, d)$ is known.

These equations apply for continuous matter density and $\vlos$ fields.
In our case, all quantities are defined on discrete grids. 
Instead of a PPD-space matter density $\rho(\glon, \glat, d)$, for example, we have a PPD 3-cube whose entries are total masses of protons $m(x, y, i)$ inside a voxel centered on $\glon_x$, $\glat_y$, and $d_i$.

To move from the continuous case to the discrete case, we assume that $\vlos(\glon, \glat, d)$ and $\sigma_v(\glon, \glat, d)$ are constant across a voxel and work with integrated quantities instead of densities. 
Equation \ref{eqn:ppdv_cont} becomes
\begin{equation}
  \label{eqn:ppdv_disc}
  \begin{split}
  &m(x, y, i, j) = m(x, y, i) \, \times \\
  &\int_{v_j - \Delta v}^{v_j + \Delta v} 
 \frac{1}{\sqrt{2 \pi \sigma_v(x, y, j)^2}} \,
    \exp\left(- \frac{\left( v - \vlos(x, y, j) \right)^2}{2 \sigma_v(x, y, j)^2} \right)\, {\rm d}v,
\end{split}
\end{equation}
where $2 \Delta v$ is the length of a voxel along the velocity axis.
Equation \ref{eqn:ppv_cont} becomes
\begin{equation}
  \label{eqn:ppv_disc}
  m(x, y, j) = \sum_{i} m(x, y, i, j).
\end{equation}
A cartoon representation of this procedure is shown in the top panel of Figure \ref{fig:diagram}.
In the cartoon, $\vlos(x, y, i)$ is assumed to be set by a rotation curve; the resulting $m(x, y, j)$ does not match the cartoon PPV data.
This mismatch is also generally true of the actual data.
Assuming a rotation curve and propagating it through produces a model PPV 3-cube which is clearly inconsistent with the observed PPV 3-cube. 

We quantify the discrepancy between a model PPV 3-cube, $m(x, y, j)$, and the observed PPV 3-cube, $PPV(x, y, j)$, with the objective function
\begin{equation}
  \label{eqn:obj_unreg}
  \begin{split}
  \mathcal{L}_u &\left(\vlos(x, y, i), \sigma_v(x, y, i) \right) = \\
 &\frac{1}{2} \sum_{x, y, j} \left(PPV(x, y, j) -  m(x, y, j)\right)^2,
 \end{split}
\end{equation}
the sum of square differences between the model and the observations. 
This is the \emph{unregularized} objective function; hence the subscript $u$ in $\mathcal{L}_u$.

We optimize $\mathcal{L}_u$ by varying the entries of $\vlos(x, y, i)$ and $\sigma_v(x, y, i)$. 
We restrict $\sigma_v(x, y, i)$ to be between $1$ and $15\text{ km sec}^{-1}$. 
The lower bound on $\sigma_v(x, y, i)$
is set to be slightly smaller than $2 \Delta v$, which in this case is $1.4 \text{ km sec}^{-1}$, as there is very little difference between pixel-convolved Gaussians with standard deviations smaller than $\Delta v$. 
We restrict $\vlos(x, y, i)$ to be within $45 \text{ km sec}^{-1}$ of the line-of-sight velocity at $\glon_x$, $\glat_y$, and $d_i$ corresponding to the IAU-recommended $220 \text{ km sec}^{-1}$ flat rotation curve and an $8.5 \text{ kpc}$ separation between the Sun and the Galactic center.
The middle panel of Figure \ref{fig:diagram} shows a cartoon solution to the the $\mathcal{L}_u$ optimization problem. 
Qualitative features of the cartoon such as the close match between the model and observed PPV 3-cubes, the magnitudes of deviations from the rotation curve, and the spatial coherence of deviations from the rotation curve are also typical of the actual solution to the unregularized problem.

This solution is not necessarily unique. 
While we have reduced the number of parameters in the problem by making assumptions about the structure of the ISM in PPDV space, there are still situations in which different $\vlos(x, y, i)$ 3-cubes produce equivalent $m(x, y, j)$ 3-cubes. 
For example, if $m(x_0, y_0, i_0) = m(x_0, y_0, i_1)$, then $\vlos(x_0, y_0, i_0)$ and $\vlos(x_0, y_0, i_1)$ are interchangeable.

To deal with this problem, we introduce external information.
The $\glon$ and $\glat$ extent of the voxels in the 3-cubes and 4-cubes are often smaller than coherent structures such as molecular clouds. 
We may therefore expect voxels that share an $\glon$ or $\glat$ boundary to have similar line-of-sight velocities.
We encode this expectation into a regularized objective function, $\mathcal{L}_r$, by adding a term penalizing large differences between the line-of-sight velocities of voxels with shared $\glon$ or $\glat$ boundaries:

\begin{equation}
  \label{eqn:obj_reg}
  \begin{split}
  \mathcal{L}_r & \left(\vlos(x, y, i), \sigma_v(x, y, i) \right) = \frac{1}{\sigma_u^2} \,
  \mathcal{L}_u  \left(\vlos(x, y, i), \sigma_v(x, y, i) \right) + \\
  & \frac{1}{2 \sigma_r^2} \sum_{x, y} \left( \vlos(x + 1, y, i) - \vlos(x, y, i) \right)^2 +\\
  & \frac{1}{2 \sigma_r^2} \sum_{x, y} \left( \vlos(x, y + 1, i) - \vlos(x, y, i) \right)^2,
  \end{split}
\end{equation}
where $\sigma_u$ and $\sigma_r$ are parameters that set the relative strengths of the model-observation residuals and regularization terms in driving the solution. 
We set $\sigma_u$ to a value corresponding to 0.05 magnitudes of reddening. This is the standard deviation reported by GSF for the distribution of residuals between the GSF reddening 3-cube integrated along the distance axis and a Planck $\tau_{353\text{GHz}}$-based integrated reddening map. 
We set $\sigma_r$ to 5 km/s.
This is approximately the value of the cloud-cloud dispersion found by \citet{Clemens:1985dp}.

The bottom panel of Figure \ref{fig:diagram} shows a cartoon representation of a regularized solution. 
The regularization is represented by springs. 
As before, the qualitative features of the cartoon are realistic. 

We evaluate the accuracy of the unregularized and regularized KT-derived $\vlos$ 3-cubes in the next section. 
For clarity, we will refer to the discretized 3-cubes as $\vlos(\glon, \glat, d)$.

\begin{figure*}[t]
\begin{center}
\includegraphics[width=1\linewidth]{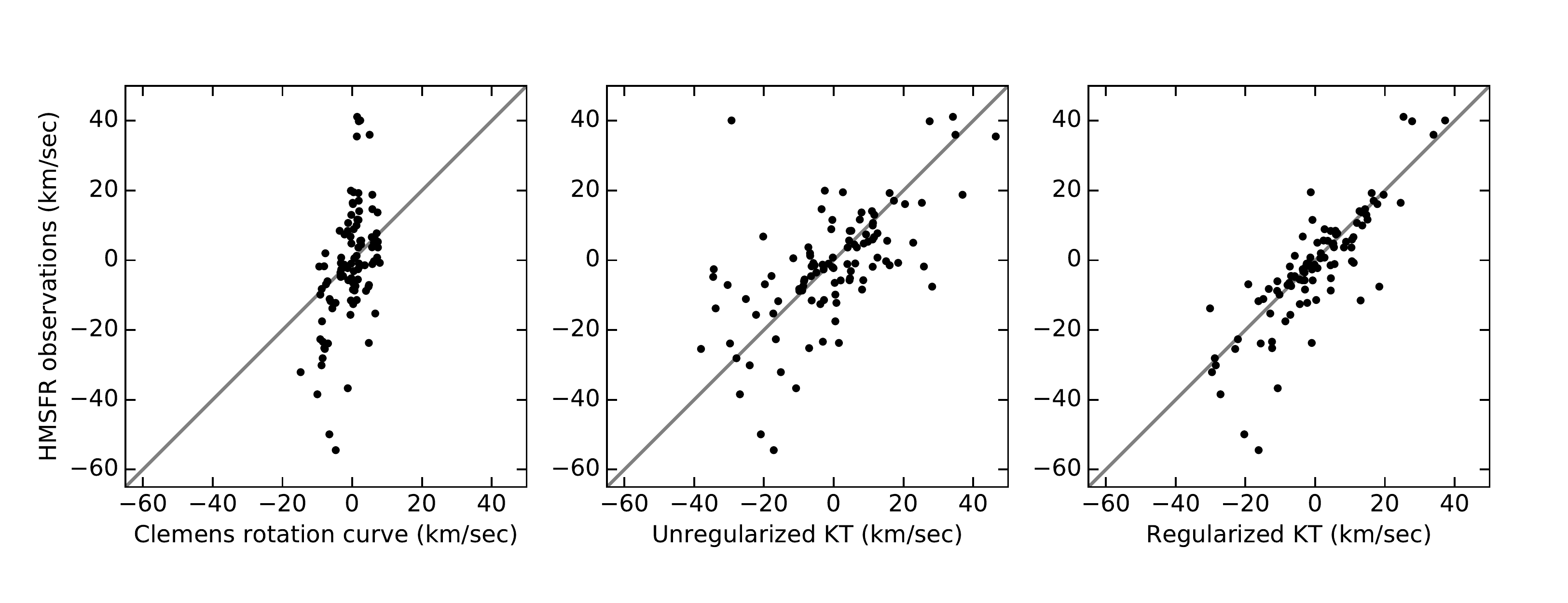}
\caption{{\label{fig:hmsfr_comparison}
The three panels show the residuals from flat rotation of the \citet{2014ApJ...783..130R} HMSFR line-of-sight velocity observations on the y-axis and residuals from flat rotation of the values of the \citet{Clemens:1985dp} rotation curve (left panel), the unregularized kinetic tomography solution (middle panel), and the regularized kinetic tomography solution (right panel) at the positions of the HMSFRs on the x-axis. In each panel, the one-to-one line is indicated in gray.
}}
\end{center}
\end{figure*}

\subsection{Checking the Kinetic Tomography solution}
\label{sec:KT-validation}
The \Reid{} HMSFRs (see \S \ref{sec:data-HMSFR}) are embedded in, born from, and, presumably, moving with dense molecular gas.
The $\vlos$ of an HMSFR should therefore be similar to the $\vlos$ of the ISM at the HMSFR's location in PPD space.
We can use this property to check the accuracy of our KT-derived $\vlos(\glon, \glat, d)$ 3-cubes by comparing the HMSFRs' observed $\vlos$ to the $\vlos$ KT assigns to the HMSFRs' $\glon$, $\glat$, and $d$ values. 
In this section, we make this comparison for the $\vlos(\glon, \glat, d)$ fields associated with a radially-varying rotation curve from \citet{Clemens:1985dp}, unregularized KT, and regularized KT. 

We must propagate the uncertainty on the distance to an HMSFR when comparing the HMSFR's observed $\vlos$ to a $\vlos(\glon, \glat, d)$ field.
The typical uncertainty on the parallax of a \Reid{} HMSFR is between $5$ and $10\%$ and is implicitly assumed to be Gaussian. 

Following the discussion in \citet{2009ApJ...704.1704B}, we assume that this Gaussian parallax uncertainty propagates linearly to a Gaussian distance uncertainty.

We assume that the uncertainty on the assigned $\vlos(\glon, \glat, d)$ value is also approximately Gaussian.
Consider an HMSFR, $s$. 
To compute the mean $\mu_s$ and standard deviation $\sigma_s$ of this HMSFR's distribution of possible $\vlos(\glon, \glat, d)$ values, we start by drawing possible distance values $d_t$ from $p_s(d)$.
Here, $p_s(d)$ is the distribution over possible distances to HMSFR $s$ and $t$ is an index over draws.
For each draw, we extract $\vlos(\glon=\glon_s, \glat=\glat_s, d=d_t)$ from $\vlos(\glon, \glat, d)$.
The mean and standard deviation of these extracted line-of-sight velocity values are $\mu_s$ and $\sigma_s$.

In Figure \ref{fig:hmsfr_comparison}, we show a comparison of the HMSFRs' observed $\vlos$ values and the mean $\vlos$ values extracted from line-of-sight velocity fields corresponding to the \citet{Clemens:1985dp} rotation curve, unregularized KT, and regularized KT.
To highlight the peculiar motions of the HMSFRs, these values are shown with the line-of-sight velocity corresponding to a flat 220 km/sec rotation and a Sun-Galactic center separation of 8.5 kpc subtracted off. 
The similarity of a velocity field to the HMSFR observations is indicated by how close points derived from that velocity field tend to be to the one-to-one line in the appropriate panel of Figure \ref{fig:hmsfr_comparison}.
It should be clear from visual inspection that the radially-varying rotation curve does not accurately predict the velocities of the HMSFRs and that both versions of KT are clearly more accurate.
There is also a significant, though possibly less visually obvious, improvement from unregularized to regularized KT.

To get a quantitative estimate of this improvement, we can compute the reduced $\chi^2$ values of the two sets of velocity estimates. 
The reduced $\chi^2$ value is given by the expression
\begin{equation}
\chi^2 = \frac{1}{\nu} \sum_s^{S} \left(\frac{{\vlos}_s - \mu_s}{\sigma_s} \right)^2, 
\end{equation}
where $\nu$ is the number of degrees of freedom in the problem and $S$ is the number of observations.
If the uncertainties are Gaussian and correctly estimated, the reduced $\chi^2$ value should be approximately equal to 1.
If we assume the regularization parameter counts against the number of degrees of freedom, the reduced $\chi^2$ values of the unregularized and regularized KT solutions are 5 and 3, respectively.

The $\chi^2$ value of the regularized KT solution is driven by 5 catastrophic outliers. 
If we remove these outliers, the reduced $\chi^2$ values of the unregularized and regularized KT solutions drop to 4 and 1.3, respectively. 
We consider the advantage of regularized over unregularized KT to be sufficient to adopt the regularized KT solution as \emph{the} KT solution, and will refer to it as such below.

At the positions of 94 of 99 HMSFRs, KT performs remarkably well.

\begin{figure*}[t]
\begin{center}
\includegraphics[width=1\linewidth]{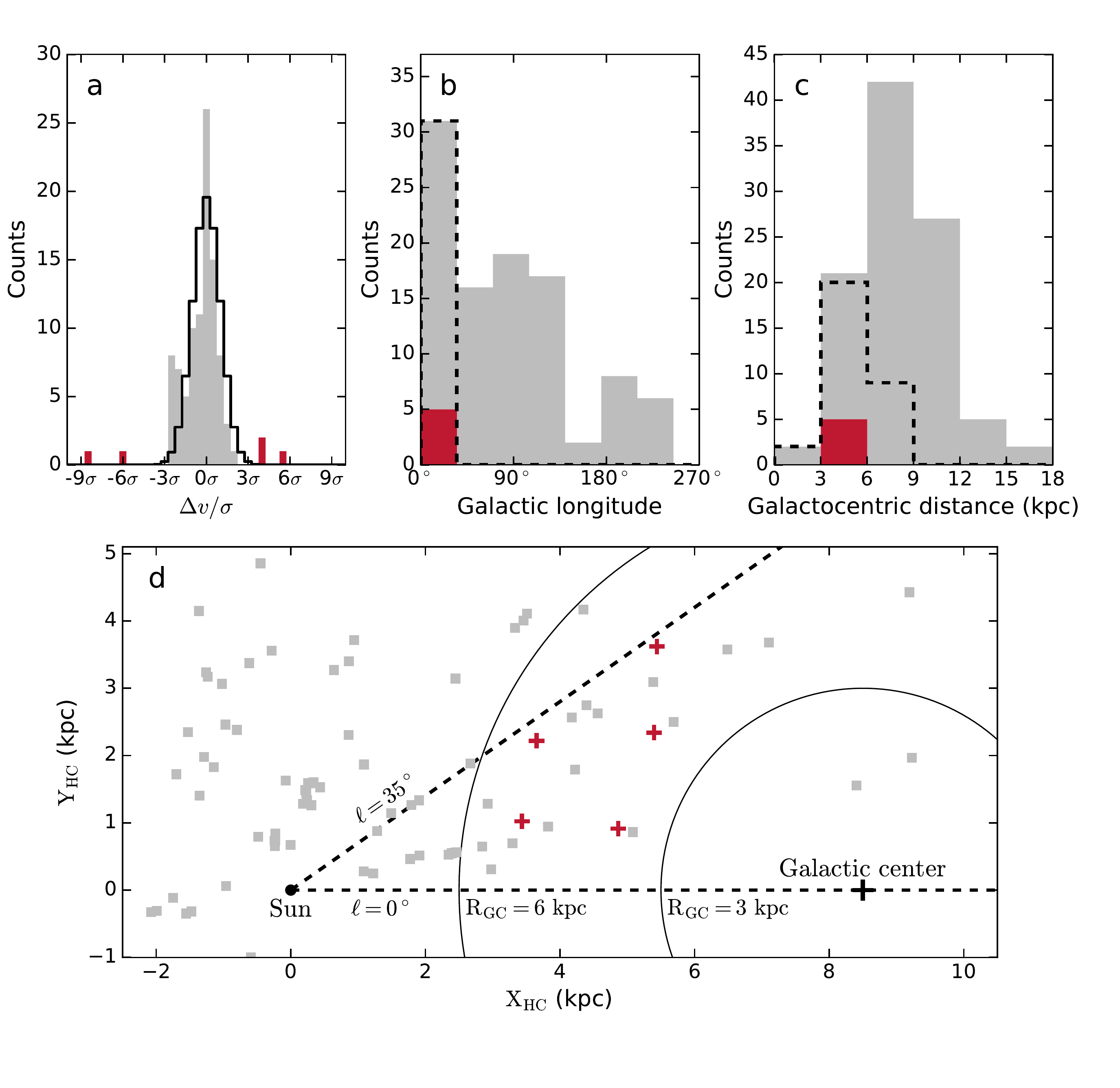}
\caption{{\label{fig:outliers}
\textbf{(a)} The distribution of residuals per standard deviation (\textit{standardized} residuals) between the observed and KT-derived line-of-sight velocities of HMSFRs (filled histogram), with an appropriately-scaled standard normal distribution for reference (empty histogram). We consider HMSFRs with $\vert \Delta v / \sigma \vert > 3$ to be outliers. In all three panels, outliers are denoted in red and non-outliers are denoted in gray.
\textbf{(b)} The distribution of the HMSFRs' Galactic longitudes. HMSFRs with $\glon < 35^\circ$ are outlined with a dashed black line.
\textbf{(c)} The distribution of the HMSFRs' Galactocentric radii. HMSFRs with $\glon < 35^\circ$ are outlined with a dashed black line.
\textbf{(d)} The locations of the HMSFRs in heliocentric Cartesian coordinates, where $\mathrm{X_{HC}}$ increases towards $\glon=0^\circ$ and $\mathrm{Y_{HC}}$ increases towards $\glon=90^\circ$.
}}
\end{center}
\end{figure*}

\subsection{Catastrophic outliers}
\label{sec:discussion-catastrophic}
In this section, we discuss the five HMSFRs where the KT-derived and directly observed $\vlos$ values are clearly inconsistent.
The distribution of standardized residuals of the HMSFR sample is shown in panel (a) of Figure \ref{fig:outliers}; the five outlier HMSFRs are marked in red.
The reason why these five HMSFRs, specifically, are outliers seems to be that they are located in a particularly complex and confused part of the Galaxy.
From panels (b), (c), and (d) of the same Figure, one can see that all of the outliers are in the inner Galaxy and relatively close to the Galactic center, at $0^\circ < \glon < 35^\circ$ and $\RGC < 6\text{ kpc}$. 
We note that four of these five outliers are in the area of sky covered by the GASS Galatic HI survey. 
We doubt GASS is to blame as it has relatively mild systematic errors and is intermediate in spatial resolution among the three surveys we use.
If we compare the distributions of all HMSFRs and outlier HMSFRs in $\glon$ and $\RGC$, it should be apparent that the distribution of outliers does not follow the distribution of all HMSFRs.

We can quantify this deviation by assuming that any HMSFR is equally likely to be one of the five outliers and computing the probability of finding all five in a randomly chosen subsample of the same size as one of our two selections.
Since a single HMSFR can not be present in a sample twice, the appropriate distribution to use for this calculation is the hypergeometric distribution, which assumes that the subsample is chosen without replacement.
The probability of finding all five outliers in a random subsample the size of either of the two selections is, in both cases, much less than 1\%.

The implication of this difference in distribution is that there is an actual enhancement in the rate at which KT produces incorrect solutions for sightlines that pass near the Galactic center. 
There are three main reasons why KT would be less accurate along a sightline at low $\glon$.
Firstly, these sightlines pass through more matter, due to both the increase in ISM surface density at low $\RGC$ and the fact that geometrically, a sightline at low $\glon$ will pass through more of the far side of the Galaxy.
Having more matter along a sightline increases the complexity of the problem --- more parcels of ISM in PPD space have to be correctly associated with velocity components in PPV space.
Secondly, due in part to the increase in ISM surface density with decreasing $\RGC$, the well-measured part of the PPD cube does not extend as far towards the inner Galaxy as it does towards the outer Galaxy (GSF). 
If the PPD cube is not accurate past some distance, then we are unlikely to obtain a correct solution past that distance.
Thirdly, the Galactic bar can stir interstellar matter up, inducing a complicated velocity field with lots of structure on small spatial scales. 
Given that the distance extent of PPD voxels near the inner Galaxy is of order a kpc, we could simply have insufficient spatial resolution to map this bar-induced velocity field. 

Two of these three reasons are caused by being at low $\glon$ while the other is caused by being in a specific $\RGC$ range. 
To try to differentiate between the two possible causes, we can repeat the quantitative exercise described above with the $\RGC$ selection as the subsample and the $\glon$ selection, rather than the full set of 99 HMSFRs, as the population. 
The probability of finding all five outliers in a random 21-element subsample of the 31-element $\glon$ selection is about 11\%. 
This probability is low but not unreasonable; we cannot differentiate between the two causes in this way.

\begin{figure*}[t]
\begin{center}
\includegraphics[width=1\linewidth]{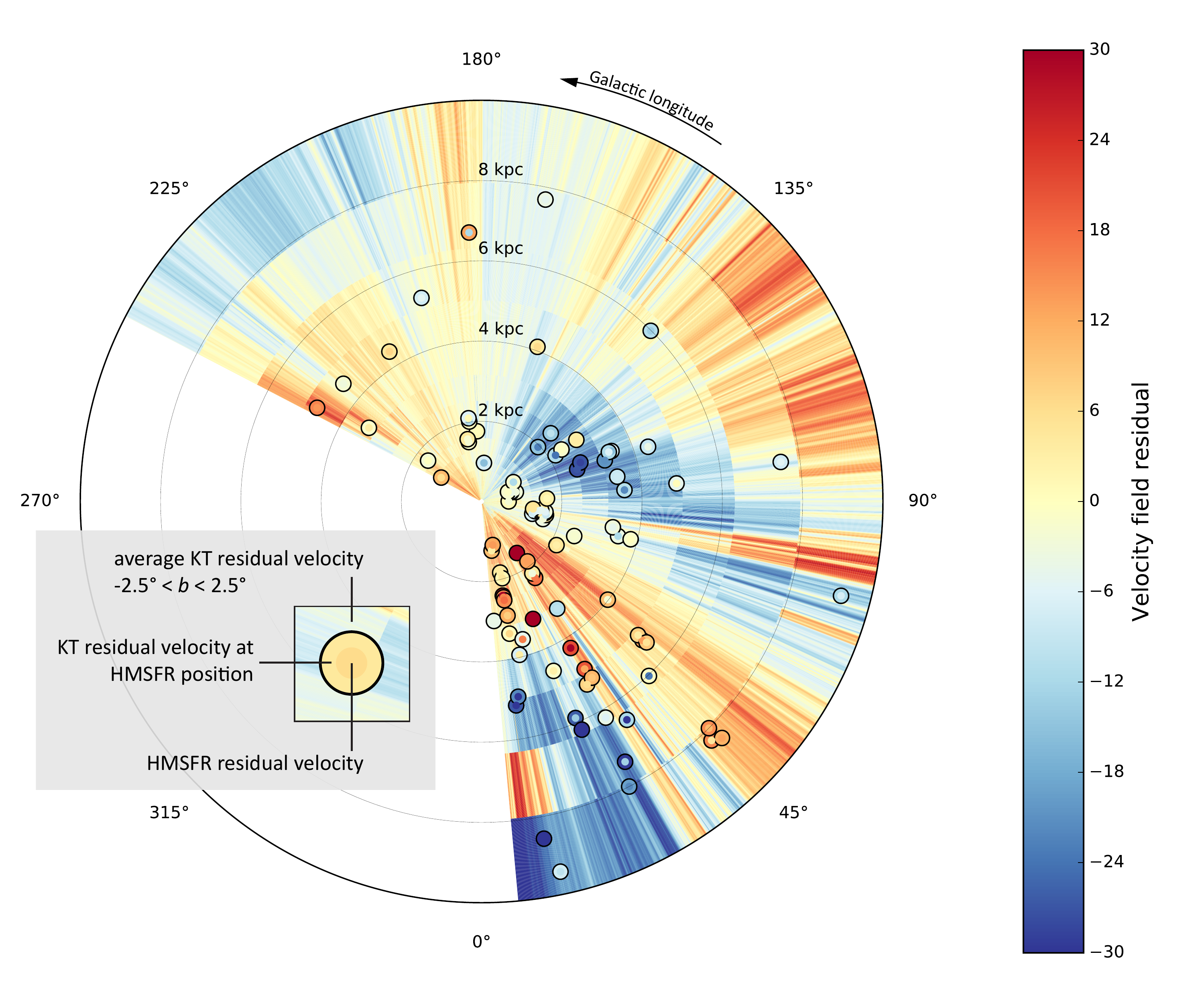}
\caption{{\label{fig:maser_pie}
Colors show the difference between an observed or estimated $\vlos$ and the value predicted from a flat rotation curve. The background is an average of the kinetic tomography-derived velocity field over $-2.5\deg \leq \glat \leq +2.5\deg$. The color of the inner part of each circle is the $\vlos$ of an HMSFR. The color of the outer ring of each circle is the value of the kinetic tomography-derived velocity field at the position (including $\glat$) of the HMSFR.
}}
\end{center}
\end{figure*}

\section{Discussion}
\label{sec:discussion}
\subsection{Qualitative features of the solution}
\label{sec:discussion-qualitative}
We have established that KT is, quantitatively-speaking, correct at the locations of 94 dense, star-forming knots of interstellar matter and we have found a heuristic explanation for why KT is quite incorrect at the locations of 5 other dense, star-forming knots of interstellar matter.
Given the lack of other published $\vlos(\glon, \glat, d)$ measurements, we cannot quantitatively establish that our $\vlos(\glon, \glat, d)$ 3-cube is also correct away from these 99 locations.
There is, however, the previously published two-dimensional peculiar velocity map of \citet{1993A&A...275...67B}.

A two-dimensional peculiar velocity field derived from the KT solution is shown in Figure \ref{fig:maser_pie}.
This peculiar velocity field is derived by subtracting the line-of-sight velocity field corresponding to a flat rotation curve from the $\vlos(\glon, \glat, d)$ 3-cube and taking the mass-weighted average of this peculiar velocity 3-cube along the $\glat$ axis over the range $-2.5^\circ \leq \glat \leq +2.5^\circ$.
The peculiar velocity field of \citet{1993A&A...275...67B} is shown in Figure 10 of their paper. 
Qualitatively, the two peculiar velocity fields are quite similar. 
The locations, extents, and signs of regions of coherent peculiar velocity are, for the most part, the same in both maps. 
There are, however, some differences in the magnitudes of peculiar velocities in these regions.

We can also qualitatively compare the typical scale of coherent velocity fluctuations in our two-dimensional peculiar velocity field to measurements of this scale from the literature. 
Much of the power in the peculiar velocity field of disc gas (e.g. \citealt{Clemens:1985dp}) and disc stars (e.g. \citealt{2015ApJ...800...83B}) is found on scales of about 2 kpc.
This is consistent, at least by eye, with the typical extent of a coherent peculiar velocity fluctuation in our two-dimensional peculiar velocity field.

Figure \ref{fig:maser_pie} also shows the line-of-sight velocities at the $\glon$, $\glat$, and $d$ of HMSFRs according to our $\vlos(\glon, \glat, d)$ 3-cube and the actual observations.
All of the HMSFRs are in $\glat$ range we averaged over to produce our two-dimensional peculiar velocity field, but there are clear differences between the observed HMSFR line-of-sight velocities and the two-dimensional velocity field. 
If we assume the two-dimensional velocity field is, for the most part, correct, then these differences imply peculiar velocity fluctuations on spatial scales of order the typical height off the plane of an HMSFR, or 10-100 pc. 
The same differences, in most cases, are also seen in the velocities assigned based on our $\vlos(\glon, \glat, d)$ 3-cube. 
Combining these facts, we can heuristically conclude that we are more-or-less correctly detecting peculiar velocity fluctuations on both large, multi-kpc scales and small, 100 pc scales. 

\subsection{Unmet assumptions and their potential consequences}
\label{sec:discussion-systematics}
Our treatment of the data, our parametrization of the 4-dimensional structure of the ISM, and our regularization scheme are all based on assumptions that are not always met.
We convert our distributions of reddening, $\atomH$ emission, and $\CO$ emission to distributions of absolute matter content in PPV and PPD space assuming there is a single, linear function relating the amount of each tracer to an amount of matter.
Due to variations in the dust-to-gas and \CO-to-\molH ratios and radiative transfer effects such as self-absorption, these functions are neither uniqiuely defined nor linear. 
We have implicitly assumed that the PPD and PPV 3-cubes are projections of the same part of the same PPDV 4-cube. 
This is not the case --- the PPV 3-cube is an integral of the PPDV 4-cube to an effectively infinite distance while the most distant well-measured voxels in the PPD 3-cube are only $\sim 10 \text{ kpc}$ away from the Sun (GSF).
Our parametrization of the 4-dimensional structure of the ISM assumes that the ISM's velocity distribution within a PPD voxel can be described by a mean, which is close to the value of the Galactic rotation curve at the voxel's center, and a dispersion.  
This will not be true for a voxel that contains a shock or is large compared to the spatial scale of velocity fluctuations; because our distance resolution is constant in log space, this is effect will apply to all voxels beyond some distance. 
Our regularization scheme assumes that the velocity distributions of voxels with shared $\glon$ or $\glat$ boundaries will be similar. 
This will once again not be true of voxels that cross shocks or are sufficiently large. 

Despite the fact that its underlying assumptions do not hold over some of the solution domain, KT produces a $\vlos(\glon, \glat, d)$ solution that quantitatively agrees with independent $\vlos(\glon, \glat, d)$ observations (see \S \ref{sec:KT-validation}).
We can resolve this conflict by concluding that (1) the assumptions, as stated, are too strict or by arguing that (2) the particular set of $\vlos(\glon, \glat, d)$ observations are not representative of the ISM as a whole.

The fact that all of the poorly-reproduced HMSFRs lie in the inner Galaxy (see \S \ref{sec:discussion-catastrophic}), where our assumptions tend to not be met more often and more egregiously than in the outer Galaxy, lends credence to the resolution (1). 
Perhaps KT is robust to some level of its assumptions not being met and this level is only exceeded in some parts of the inner Galaxy. 
As an example of the extent to which the inner Galaxy does not meet our assumptions, consider the fact that the maximum distance from the Sun to which the PPD 3-cube is accurate to in the inner Galaxy is approximately 5 kpc GSF while the PPV 3-cube is integrated out to the far edge of the Galaxy. 

Resolution (2), that the comparison observations are atypical, seems unlikely considering the qualitative structure of the solution (see \S \ref{sec:discussion-qualitative}) but cannot yet be ruled out empirically. 
The comparison observations are of HMSFRs, which by definition will be associated with large overdensities of molecular gas. 
It is possible that the check we have performed in \S \ref{sec:KT-validation} only applies to dense molecular gas.
For example, KT could merely be finding the most massive object along a sightline through the PPD 3-cube and associating it with the most massive objective in the corresponding sightline through the PPV 3-cube. 
This would usually correctly assign a distance and velocity to something like a GMC but would fail for less concentrated neutral gas. 
However, it would be difficult to reconcile the clear and non-trivial larger-scale velocity structure seen in the previous section with a scenario in which the KT solution is only correct at extreme overdensities. 

To get a more quantitative and broadly applicable understanding of when KT works and fails, we would need either a less density-biased set of independent $\vlos(\glon, \glat, d)$ measurements or a set of artificial injection tests. 
These artificial injection tests would consist of numerical experiments in which we artificially observe a model galaxy's ISM, reconstruct the $\vlos$ field from these artificial observations using KT, and compare the reconstructed and input model $\vlos$ fields. 
To the best of our knowledge, there are no currently available catalogs of these sorts of less density-biased measurements, ruling out option one.
The steps involved in artificial injection tests, particularly simulating galaxies at sufficiently high resolution and producing artificial observations in a way that includes the non-trivial systematics in the actual observations, are complicated enough to put option two beyond the scope of this work.

Both of these options are plausible directions for future work. 
The 1.527 $\mu$m diffuse interstellar band (DIB), for instance, has been mapped over much of the northern sky by the APOGEE survey \citep{2015ApJ...798...35Z}. 
Observations of this DIB towards APOGEE stars with known distances could potentially be used to build catalogs or even maps of $\vlos(\glon, \glat, d)$ using an independent dataset. 
Artificial injection tests are conceptually straightforward, though they do require a substantial investment of time and computational resources.
While neither option is easy, we would argue that some combination of the two will be necessary before the KT-derived $\vlos$ 3-cube can be trusted away from the HMSFRs of \Reid{}.

\section{Conclusion}
\label{sec:conclusion}

In this work we developed a method for measuring the radial velocity of parcels of the interstellar medium of a measured distance, which we dubbed  Kinetic Tomography. We argued that this method is important as a tool for measuring converging and diverging flows around our Galaxy as well as for detecting large scale deviations from assumed rotation curves. The method takes as inputs the three-dimensional distribution of dust in our Galaxy measured from stellar photometry and the emission spectra from Galatic $\CO$ and $\atomH$. We developed a technique that assigns each 3D parcel of ISM from the dust map a line-of-sight velocity and line-of-sight velocity width, in order to best reproduce the observed $\CO$ and $\atomH$ data. We found that we can improve the fidelity of our solution by implementing Tikhonov regularization, effectively coupling the line-of-sight velocity of adjacent pixels. 

As a test of our method we compare our results to independent measurements of HMSFRs from \Reid{}, which contain both distances and line-of-sight velocity information. We find that of the 99 HMSFRs in the area of sky we study, 94 are consistent with our results and 5 are outliers, all of which lie near Galactic center. This consistency indicates our map is an accurate representation of the velocity field of the ISM, at least in denser regions consistent with HMSFRs. We also find qualitative consistency with the peculiar velocity maps of \citet{1993A&A...275...67B}. 

Here we conclude that KT can be a very powerful tool for the study of the velocity structure of the Galactic ISM. In future work, we will investigate what KT can tell us about the Galactic rotation curve, streaming motions within the Galactic disk, and the vertical structure of Galactic flows.

\section{Acknowledgements}
We thank the anonymous referee for their concise and constructive comments.
We thank Greg Green and Eddie Schlafly for guidance on the use of the GSF reddening cube.
This material is based upon work supported by the National Science Foundation under Grant No. 1616177 and the Space Telescope Science Institute Director's Discretionary Fund.

\bibliographystyle{aasjournal}

\end{document}